%% file: paper.tex
\documentclass[a4paper,envcountsame]{easychair}

\usepackage{amsfonts}
\usepackage{amsmath}
\usepackage{amssymb}
\usepackage{algorithm}
\usepackage{algorithmic}
\usepackage{graphicx}

\newtheorem{theorem}{Theorem}
\newtheorem{proposition}[theorem]{Proposition}
\newtheorem{definition}[theorem]{Definition}

\DeclareMathOperator*{\argmax}{argmax}

\DeclareMathOperator{\attr}{Attr}
\DeclareMathOperator{\ind}{Index}
\DeclareMathOperator{\pre}{Pre}
\DeclareMathOperator{\play}{Play}
\DeclareMathOperator{\br}{br}
\DeclareMathOperator{\val}{Val}
\DeclareMathOperator{\gamevalue}{Value}
\DeclareMathOperator{\all}{All}

\DeclareMathOperator{\subtree}{Subtree}
\DeclareMathOperator{\critical}{Critical}
\DeclareMathOperator{\escapes}{Esc}
\DeclareMathOperator{\fixsnare}{FixSnare}

\DeclareMathOperator{\augment}{Augment}

\DeclareMathOperator{\increase}{Increase}

\DeclareMathOperator{\snareincrease}{SnareIncrease}
\DeclareMathOperator{\snare}{Snare}
\DeclareMathOperator{\nonobliv}{NonOblivious}
\DeclareMathOperator{\policy}{Policy}
\DeclareMathOperator{\zeromean}{ZeroMeanPartition}
\DeclareMathOperator{\schewe}{Optimal}

\newcommand{\nats}{\mathbb N}
\newcommand{\ints}{\mathbb Z}
\newcommand{\meanpayoff}{\mathcal M}
\newcommand{\weight}{w}
\newcommand{\tree}{T}

\begin{document}

\input{intro.tex}

\input{definitions.tex}
\input{snareintro.tex}
\input{si.tex}

\input{trees.tex}
\input{backedges.tex}

\input{snares.tex}

\input{friedmann.tex}

\input{conclusions.tex}

\bibliographystyle{abbrv}
\bibliography{../../references/references.bib}

\appendix
\input{appendix.tex}

\end{document}

%% file: intro.tex
\title{Non-oblivious Strategy Improvement}
\author{John Fearnley \\
Department of Computer Science, University of Warwick, UK}
\authorrunning{John Fearnley}
\titlerunning{Non-oblivious Strategy Improvement}
\date{}
\maketitle

\begin{abstract}
We study strategy improvement algorithms for mean-payoff and parity games. We
describe a structural property of these games, and we show that these structures
can affect the behaviour of strategy improvement. We show how awareness of these
structures can be used to accelerate strategy improvement algorithms. We call
our algorithms non-oblivious because they remember properties of the game that
they have discovered in previous iterations. We show that non-oblivious strategy
improvement algorithms perform well on examples that are known to be hard for
oblivious strategy improvement. Hence, we argue that previous strategy
improvement algorithms fail because they ignore the structural properties of the
game that they are solving.
\end{abstract}

\section{Introduction}
In this paper we study strategy improvement for two player infinite games played
on finite graphs. In this setting the vertices of a graph are divided between
two players. A token is placed on one of the vertices, and in each step the
owner of the vertex upon which the token is placed must move the token along one
of the outgoing edges of that vertex. In this fashion, the two players form an
infinite path in the graph. The payoff of the game is then some property of this
path, which depends on the type of game that is being played.  Strategy
improvement is a technique that originated from Markov decision
processes~\cite{howard60}, and has since been applied many types of games in
this setting, including simple stochastic games~\cite{condon93},
discounted-payoff games~\cite{puri95}, mean-payoff games~\cite{bv07}, and parity
games~\cite{jurdzinski00b,bsv03}. In this paper we will focus on the strategy
improvement algorithm of Bj\"orklund and Vorobyov~\cite{bv07}, which is designed
to solve mean-payoff games, but can also be applied to parity games.

Algorithms that solve parity and mean-payoff games have received much interest.  One
reason for this is that the model checking problem for the modal $\mu$-calculus
is polynomial time equivalent to the problem of solving a parity
game~\cite{EmersonJutla93,stirling95}, and there is a polynomial time reduction
from parity games to mean-payoff games~\cite{puri95}. Therefore, faster
algorithms for these games lead to faster model checkers for the $\mu$-calculus.
Secondly, both of these games lie in NP~$\cap$~co-NP, which implies that neither
of the two problems are likely to be complete for either class.  Despite this,
no polynomial time algorithms have been found.


The approach of strategy improvement can be described as follows. The algorithm
begins by choosing one of the players to be the strategy improver, and then
picks an arbitrary strategy for that player. A strategy for a
player consists of a function that picks one edge for each of that player's
vertices. Strategy improvement then computes a set of profitable edges for that
strategy.  If the strategy is switched so that it chooses some subset of the
profitable edges, rather than the edges that are currently chosen, then strategy
improvement guarantees that the resulting strategy is better in some
well-defined measure. So, the algorithm picks some subset of the profitable
edges to create a new, improved, strategy to be considered in the next
iteration. This process is repeated until a strategy is found that has no
profitable edges, and this strategy is guaranteed optimal for the strategy
improver. Since any subset of the profitable edges could be used to create an
improved strategy in each iteration, some method is needed to determine which
subset to choose in each iteration. We call this method a switching policy, and
the choice of switching policy can have a dramatic effect on the running time of
the algorithm.

A significant amount of research has been dedicated to finding good switching
policies. In terms of complexity bounds, the current best switching policies are
randomized, and run in an expected $O(2^{\sqrt{n \log n}})$ number of
iterations~\cite{bv07}. Another interesting switching policy is the optimal
switching policy given by Schewe~\cite{schewe08}. An optimal switching policy
always picks the subset of profitable edges that yields the best possible
successor strategy, according to the measure that strategy improvement uses to
compare strategies. It is not difficult to show that such a subset of profitable
edges must exist, but computing an optimal subset of profitable edges seemed to
be difficult, since there can be exponentially many subsets of profitable edges
to check. Nevertheless, Schewe's result is a polynomial time algorithm that
computes an optimal subset of edges. Therefore, optimal switching policies can
now be realistically implemented. It is important to note that the word
``optimal'' applies only to the subset of profitable edges that is chosen to be
switched in each iteration. It is not the case that a strategy improvement
algorithm equipped with an optimal switching policy will have an optimal running
time.

Perhaps the most widely studied switching policy is the all-switches policy,
which simply selects the entire set of profitable edges in every iteration.
Although the best upper bound for this policy is $O(2^{n}/n)$
iterations~\cite{mansoursingh99}, it has been found to work extremely well in
practice. Indeed, for a period of ten years there were no known examples upon
which the all switches policy took significantly more than a linear number of
iterations. It was for this reason that the all-switches policy was widely held
to be a contender for a proof of polynomial time termination.

However, Friedmann has recently found a family of examples that force a strategy
improvement algorithm equipped with the all-switches policy to take an
exponential number of steps~\cite{Fri09}. Using the standard
reductions~\cite{puri95,patersonzwick96}, these examples can be generalised to
provide exponential lower bounds for all-switches on mean-payoff and
discounted-payoff games. Even more surprisingly, Friedmann's example can be
generalised to provide an exponential lower bound for strategy improvement
algorithms equipped with an optimal switching policy~\cite{Fri09preprint}.
This recent revelation appears to imply that there is no longer any hope for
strategy improvement, since an exponential number of iterations can be forced
even if the best possible improvement is made in every step.

\paragraph{Our contributions.} Despite ten years of research into strategy
improvement algorithms, and the recent advances in the complexity of some widely
studied switching policies, the underlying combinatorial structure of
mean-payoff and parity games remains somewhat mysterious. There is no previous
work which links the structural properties of a parity or mean-payoff
game with the behaviour of strategy improvement on those games. In this paper,
we introduce a structural property of these games that we call
a snare. We show how the existence of a snare in a parity or mean-payoff game
places a restriction on the form that a winning strategy can take for these
games. Hence, we argue that every algorithm that computes a winning strategy for
these games must, at least implicitly, deal with these structures.

In the case of strategy improvement algorithms, we argue that snares play a
fundamental role in the behaviour of these algorithms. We show that there is a
certain type of profitable edge, which we call a back edge, that is the
mechanism that strategy improvement uses to deal with snares. We show how each
profitable back edge encountered by strategy improvement corresponds to some
snare that exists in the game. Hence, we argue that the concept of a snare is
a new tool that can be used in the analysis of strategy improvement algorithms.

We then go on to show that, in addition to being an analytical tool, awareness
of snares can be used to accelerate the process of strategy improvement. We
propose that strategy improvement algorithms should remember the snares that
they have seen in previous iterations, and we give a procedure that uses a
previously recorded snare to improve a strategy. Strategy improvement algorithms
can choose to apply this procedure instead of switching a subset of profitable
edges. We give one reasonable example of a strategy improvement algorithm that
uses these techniques. We call our algorithms non-oblivious strategy improvement
algorithms because they remember information about their previous iterations,
whereas previous techniques make their decisions based only on the information
available in the current iteration. 

In order to demonstrate how non-oblivious techniques can be more powerful than
traditional strategy improvement, we study Friedmann's family of examples that
cause the all-switches and the optimal switching policies to take exponential
time. We show that in certain situations non-oblivious strategy improvement
makes better progress than even the optimal oblivious switching policy. We go on
to show that this behaviour allows our non-oblivious strategy improvement
algorithms to terminate in polynomial time on Friedmann's examples. This fact
implies that it is ignorance of snares that is a key failing of oblivious
strategy improvement.

%% file: definitions.tex
\section{Preliminaries} 

A mean-payoff game is defined by a tuple $(V, V_\text{Max}, V_\text{Min}, E,
\weight)$
where~$V$ is a set of vertices and~$E$ is a set of edges, which together form a
finite graph. Every vertex must have at least one outgoing edge.  The
sets~$V_\text{Max}$ and $V_{\text{Min}}$ partition~$V$ into vertices belonging
to player Max and vertices belonging to player Min, respectively. The function
$\weight : V \rightarrow \ints$ assigns an integer weight to every vertex. 

The game begins by placing a token on a starting vertex~$v_0$. In each step, the
player that owns the vertex upon which the token is placed must choose one
outgoing edge of that vertex and move the token along it. In this fashion, the
two players form an infinite path $\pi = \langle v_0, v_1, v_2, \dots \rangle$,
where~$(v_i, v_{i+1})$ is in~$E$ for every~$i$ in~$\nats$.  The \emph{payoff} of an
infinite path is defined to be $\meanpayoff (\pi) = \lim\inf_{n \rightarrow
\infty} (1/n) \sum_{i=0}^{n} \weight(v_i)$. The objective of Max is to maximize the
value of~$\meanpayoff (\pi)$, and the objective of Min is to minimize it.

A \emph{positional strategy} for Max is a function that chooses one outgoing edge for
every vertex belonging to Max. A strategy is denoted by $\sigma : V_\text{Max}
\rightarrow V$, with the condition that $(v, \sigma(v))$ is in $E$, for every
Max vertex~$v$. Positional strategies for player Min are defined analogously.
The sets of positional strategies for Max and Min are denoted by
$\Pi_\text{Max}$ and $\Pi_\text{Min}$, respectively.  Given two positional
strategies,~$\sigma$ and~$\tau$ for Max and Min respectively, and a starting
vertex $v_0$, there is a unique path $\langle v_0, v_1, v_2 \dots \rangle$,
where~$v_{i+1} = \sigma(v_i)$ if $v_i$ is owned by Max and~$v_{i+1} = \tau(v_i)$
if $v_i$ is owned by Min. This path is known as the \emph{play} induced by the two
strategies~$\sigma$ and~$\tau$, and will be denoted by $\play(v_0, \sigma,
\tau)$.

For all~$v$ in~$V$ we define:
\begin{align*}
\gamevalue_{*}(v) &= \max_{\sigma \in \Pi_\text{Max}} \min_{\tau \in \Pi_\text{Min}} \meanpayoff(\play(v, \sigma,
\tau))\\
\gamevalue^{*}(v) &= \min_{\tau \in \Pi_\text{Min}} \max_{\sigma \in \Pi_\text{Max}} \meanpayoff(\play(v, \sigma, \tau))
\end{align*}
These are known as the lower and upper values, respectively. For mean-payoff
games we have that the two quantities are equal, a property called determinacy. 
\begin{theorem}[\textbf{\cite{ll69}}]
For every starting vertex $v$ in every mean-payoff game we have $\gamevalue_*(v) =
\gamevalue^*(v)$.
\end{theorem}
For this reason, we define $\gamevalue(v)$ to be the value of the game starting at the
vertex $v$, which is equal to both $\gamevalue_*(v)$ and $\gamevalue^*(v)$. The
computational task associated with mean-payoff games is to find~$\gamevalue(v)$ for
every vertex~$v$. 

Computing the 0-mean partition is a decision version of this problem. This
requires us to decide whether $\gamevalue(v) > 0$, for every vertex~$v$. 
Bj\"orklund and Vorobyov have shown that only a polynomial number of calls to an
algorithm for finding the 0-mean partition are needed to find the value for
every vertex in a mean-payoff game~\cite{bv07}.

A Max strategy $\sigma$ is a \emph{winning strategy} for a set of vertices $W$ if
$\meanpayoff(v, \sigma, \tau) > 0$ for every Min strategy~$\tau$ and every
vertex $v$ in $W$. Similarly, a Min strategy~$\tau$ is a winning strategy for
$W$ if $\meanpayoff(v, \sigma, \tau) \le 0$ for every Max strategy~$\sigma$ and
every vertex $v$ in $W$. To solve the 0-mean partition problem we are required
to partition the vertices of the graph into the sets $(W_{\text{Max}},
W_\text{Min})$, where Max has a winning strategy for~$W_\text{Max}$ and Min has
a winning strategy for~$W_{\text{Min}}$.

%% file: snareintro.tex
\section{Snares}
\label{snareintro}

In this section we introduce a structure called that we call a ``snare''. The
dictionary definition\footnote{American Heritage Dictionary of the English
Language, Fourth Edition} of the word snare is ``something that serves to
entangle the unwary''. This is a particularly apt metaphor for these structures
since, as we will show, a winning strategy for a player must be careful to avoid
being trapped by the snares that are present in that player's winning set.

The definitions in this section could be formalized for either player. We choose
to focus on player Max because we will later choose Max to be the strategy
improver. For a set of vertices $W$ we define $G \restriction W$ to be the
sub-game induced by $W$, which is $G$ with every vertex not in $W$ removed. A
snare for player Max is defined to be a subgame for which player Max can
guarantee a win from every vertex.
 
\begin{definition}[Max Snare]
For a game $G$, a snare is defined to be a tuple $(W, \chi)$ where $W \subseteq
V$ and $\chi : W \cap V_\text{Max} \rightarrow W$ is a partial strategy for player Max
that is winning for every vertex in the subgame $G \restriction W$.
\end{definition}

This should be compared with the concept of a dominion that was introduced by
Jurdzi\'nski, Paterson, and Zwick~\cite{jpz06}. A dominion is also a subgame in
which one of the players can guarantee a win, but with the additional constraint
that the opponent is unable to leave the dominion. By contrast, the opponent
may be capable of leaving a snare. We define an escape edge for Min to be an
edge that Min can use to leave a Max snare.

\begin{definition}[Escapes]
Let~$W$ be a set of vertices. We define the escapes from~$W$ as $\escapes(W) =
\{(v, u) \in E \; : \; v \in W \cap V_\text{Min} \text{ and } u \notin W \}$
\end{definition}

It is in Min's interests to use at least one escape edge from a snare, since
if Min stays in a Max snare forever, then Max can use the strategy $\chi$ to
ensure a positive payoff. In fact, we can prove that if $\tau$ is a winning
strategy for Min for some subset of vertices then $\tau$ must use at least one escape
from every Max snare that exists in that subset of vertices.

\begin{theorem}
\label{winningsnares}
Suppose that $\tau$ is a winning strategy for Min on a set of vertices $S$. If
$(W, \chi)$ is a Max snare where $W \subset S$, then there is some
edge $(v, u)$ in $\escapes(W)$ such that $\tau(v) = u$. 
\end{theorem}

\begin{figure}
  \begin{center}
    \includegraphics[width=0.27\textwidth]{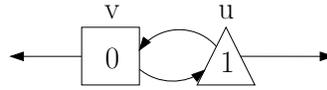}
  \end{center}
  \caption{A simple snare.}
  \label{snare-figure}
\end{figure}

Figure~\ref{snare-figure} shows an example of a subgame upon which a snare can
be defined. In all of our diagrams, boxes are used to represent Max vertices and
triangles are used to represent Min vertices. The weight assigned to each vertex
is shown on that vertex. If we take $W = \{v, u\}$ and $\chi(v) = u$ then $(W,
\chi)$ will be a Max snare in every game that contains this structure as a
subgame. This is because the cycle is positive, and therefore $\chi$ is a
winning for Max on the subgame induced by $W$. There is one escape from this
snare, which is the edge Min can use to break the cycle at $u$.

Since the example is so simple, Theorem~\ref{winningsnares} gives a particularly
strong property for this snare: every winning strategy for Min must use the
escape edge at $u$. If Min uses the edge $(u, v)$ in some strategy, then Max can
respond by using the edge $(v, u)$ to guarantee a positive cycle, and therefore
the strategy would not be winning for Min. This is a strong property because we
can essentially ignore the edge $(u, v)$ in every game into which the example is
embedded. This property does not hold for snares that have more than one escape.



%% file: si.tex
\section{Strategy Improvement}

In this section we will summarise Bj\"orklund and Vorobyov's strategy
improvement algorithm for finding the 0-mean partition of a mean-payoff
game~\cite{bv07}. Their algorithm requires that the game is modified by adding
retreat edges from every Max vertex to a special sink vertex.
\begin{definition}[Modified Game]
A game $(V, V_\text{Max}, V_\text{Min}, E, \weight)$ will be modified to create $(V \cup \{s\}, V_\text{Max} \cup \{s\}, V_\text{Min},
E', \weight')$, where $E' = E \cup \{(v, s) \; : \; v \in V_\text{Max} \}$, and
$\weight'(v) = \weight(v)$ for all vertices $v$ in $V$, and $\weight'(s) = 0$.
\end{definition}
Strategy improvement always works with the modified game, and for the rest of
the paper we will assume that the game has been modified. 

Given two strategies, one for each player, the play induced by the two
strategies is either a finite path that ends at the sink or a finite initial
path followed by an infinitely repeated cycle. This is used to define the
valuation of a vertex. 

\begin{definition}[Valuation]
Let $\sigma$ be a positional strategy for Max and $\tau$ be a positional
strategy for Min. If $\play(v_0, \sigma, \tau) = \langle v_0, v_1, \dots v_k,
\langle c_0, c_1, \dots c_l \rangle^{\omega} \rangle$, for some vertex $v_0$,
then we define $\val^{\sigma,\tau}(v_0) = -\infty$ if $\sum_{i=0}^l
\weight(c_i) \leq 0$ and $\infty$ otherwise. Alternatively, if $\play(v, \sigma,
\tau) = \langle v_0, v_1, \dots v_k, s \rangle$ then we define $\val^{\sigma,
\tau}(v_0) = \sum_{i=0}^{k} \weight(v_i)$.
\end{definition}

Strategy improvement algorithms choose one player to be the strategy improver,
which we choose to be Max. For a Max strategy~$\sigma$, we define~$\br(\sigma)$
to be the \emph{best response} to~$\sigma$, which is a Min strategy with the
property $\val^{\sigma, \br(\sigma)}(v) \leq \val^{\sigma, \tau}(v)$ for every
vertex~$v$ and every Min strategy~$\tau$. Such a strategy always exists, and
Bj\"orklund and Vorobyov give a method to compute it in polynomial
time~\cite{bv07}.  We will frequently want to refer to the valuation of a vertex
$v$ when the Max strategy~$\sigma$ is played against~$\br(\sigma)$, so we
define~$\val^{\sigma}(v)$ to be shorthand for $\val^{\sigma, \br(\sigma)}(v)$.
Occasionally, we will need to refer to valuations from multiple games. We use
$\val^{\sigma}_G(v)$ to give the valuation of the vertex~$v$ when $\sigma$ is
played against $\br(\sigma)$ in the game~$G$. We extend all of our notations in
a similar manner, by placing the game in the subscript.

For a Max strategy~$\sigma$ and an edge~$(v, u)$ that is not chosen by~$\sigma$,
we say~$(v, u)$ is \emph{profitable} in~$\sigma$ if $\val^\sigma(\sigma(v))
< \val^{\sigma}(u)$. \emph{Switching} an edge~$(v, u)$ in~$\sigma$ is denoted by
$\sigma[v \mapsto u]$. This operation creates a new strategy where, for a vertex
$w \in V_{\text{Max}}$ we have $\sigma[v \mapsto u](w) = u$ if $w = v$, and
$\sigma(w)$ otherwise. Let $F$ be a set of edges that contains at most one
outgoing edge from each vertex. We define~$\sigma[F]$ to be~$\sigma$ with every
edge in~$F$ switched. The concept of profitability is important because 
switching profitable edges creates an improved strategy.
\begin{theorem}[\cite{bv07}]
\label{bv1}
Let~$\sigma$ be a strategy and~$P$ be the set of edges that are profitable
in~$\sigma$. Let $F \subseteq P$ be a subset of the profitable edges that
contains at most one outgoing edge from each vertex. For every vertex~$v$ we
have $\val^{\sigma}(v) \leq \val^{\sigma[W]}(v)$, and there is a vertex for
which the inequality is strict.
\end{theorem}

The second property that can be shown is that a strategy with no profitable
edges is optimal. An optimal strategy is a Max strategy~$\sigma$ such that
$\val^{\sigma}(v) \ge \val^{\chi}(v)$ for every Max strategy~$\chi$ and every
vertex $v$. The 0-mean partition can be derived from an optimal
strategy~$\sigma$: the set $W_\text{Max}$ contains
every vertex~$v$ with $\val^{\sigma}(v) = \infty$, and $W_\text{Min}$ contains
every vertex $v$ with $\val^{\sigma}(v) < \infty$.

\begin{theorem}[\cite{bv07}]
\label{bv2}
A strategy with no profitable edges is optimal.
\end{theorem}

Strategy improvement begins by choosing a strategy $\sigma_0$ with the property
that $\val^{\sigma_0}(v) > -\infty$ for every vertex $v$. One way to achieve
this is to set $\sigma_0(v) = s$ for every vertex $v$ in $V_\text{Max}$.  This
guarantees the property unless there is some negative cycle that Min can enforce
without passing through a Max vertex. Clearly, for a vertex $v$ on one
of these cycles, Max has no strategy $\sigma$ with
$\val^{\sigma}(v) > -\infty$. These vertices can therefore be removed in a
preprocessing step and placed in $W_\text{Min}$.

For every strategy $\sigma_i$ a new strategy $\sigma_{i+1} = \sigma_i[F]$ will
be computed, where~$F$ is a subset of the profitable edges in $\sigma_i$, which
contains at most one outgoing edge from each vertex.  Theorem~\ref{bv1} implies
that $\val^{\sigma_{i+1}}(v) \ge \val^{\sigma_i}(v)$ for every vertex $v$, and
that there is a vertex for which the inequality is strict.  This implies that a
strategy cannot be visited twice by strategy improvement. The fact that there is
a finite number of positional strategies for Max implies that strategy
improvement must eventually reach a strategy $\sigma_k$ in which no edges are
profitable. Theorem~\ref{bv2} implies that $\sigma_k$ is the optimal strategy,
and strategy improvement terminates.

Strategy improvement requires a rule that determines which profitable edges are
switched in each iteration. We will call this a \emph{switching policy}. Oblivious
switching policies are defined as $\alpha : 2^E \rightarrow 2^E$, where for
every set~$P \subseteq E$, we have that~$\alpha(P)$ contains at most one
outgoing edge for each vertex.

Some of the most widely studied switching policies are all-switches policies.
These policies always switch every vertex that has a profitable edge, and when a
vertex has more than one profitable edge an additional rule must be given to
determine which edge to choose.  Traditionally this choice is made by choosing
the successor with the highest valuation. We must also be careful to break ties
when there are two or more successors with the highest valuation. Therefore, for
the purposes of defining this switching policy we will assume that each vertex
$v$ is given a unique index in the range $\{1, 2, \dots, |V|\}$, which we will
denote as $\ind(v)$.
\begin{equation*}
\begin{split}
\all(F) = \{ (v, u) \; : \; \text{There is no edge } (v, w) \in F \text{ with }
\val^{\sigma}(u) < \val^{\sigma}(w) \\ 
\text{ or with } \val^{\sigma}(u) = \val^{\sigma}(w) \text{ and } \ind(u) <
\ind(w)\}.
\end{split}
\end{equation*}

In the introduction we described optimal switching policies, which we can now
formally define.  A switching policy is optimal if it selects a subset of
profitable edges~$F$ that satisfies $\val^{\sigma[H]}(v) \leq
\val^{\sigma[F]}(v)$ for every subset of profitable edges~$H$ and every
vertex~$v$. Schewe has given a method to compute such a set in polynomial
time~\cite{schewe08}. We will denote an optimal switching policy as $\schewe$.

%% file: trees.tex
\section{Strategy Trees}
\label{strategytrees}

The purpose of this section is to show how a strategy and its best response can
be viewed as a tree, and to classify profitable edges by their position in this
tree. We will classify edges as either cross edges or back edges. We will later
show how profitable back edges are closely related to snares.
 
It is technically convenient for us to make the assumption that every vertex has
a finite valuation under every strategy. The choice of starting strategy ensures
that for every strategy $\sigma$ considered by strategy improvement, we have
$\val^{\sigma}(v) > -\infty$ for every vertex $v$. Obviously, there may be
strategies under which some vertices have a valuation of $\infty$. The first
part of this section is dedicated to rephrasing the problem so that our
assumption can be made.

We define the \emph{positive cycle} problem to be the problem of finding a
strategy~$\sigma$ with $\val^{\sigma}(v) = \infty$ for some vertex $v$, or to
prove that there is no strategy with this property. The latter can be done by
finding an optimal strategy $\sigma$ with $\val^{\sigma}(v) < \infty$ for every
vertex~$v$. We can prove that a strategy improvement algorithm for the positive
cycle problem can be adapted to find the 0-mean partition.

\begin{proposition}
\label{poscycle}
Let $\alpha$ be a strategy improvement algorithm that solves the positive cycle
problem in $O(\kappa)$ time. There is a strategy improvement algorithm which
finds the 0-mean partition in $O(|V| \cdot \kappa)$ time.
\end{proposition}

We consider switching policies that solve the positive cycle problem, and so we
can assume that every vertex has a finite valuation under every strategy that
our algorithms consider. Our switching policies will terminate when a vertex
with infinite valuation is found.  With this assumption we can define the
strategy tree.
\begin{definition}[Strategy Tree]
Given a Max strategy $\sigma$ and a Min strategy $\tau$ we define the tree
$\tree^{\sigma, \tau} = (V, E')$ where $E' = \{ (v, u) \; : \; \sigma(v) = u
\text{ or } \tau(v) = u \}$.
\end{definition}
In other words, $\tree^{\sigma, \tau}$ is a tree rooted at the sink whose edges
are those chosen by $\sigma$ and $\tau$. We define $\tree^{\sigma}$ to be
shorthand for $\tree^{\sigma, \br(\sigma)}$, and $\subtree^{\sigma}(v): V
\rightarrow 2^V$ to be the function that gives the vertices in the subtree
rooted at the vertex $v$ in $T^{\sigma}$.

\begin{figure}
  \begin{center}
    \includegraphics[width=0.35\textwidth]{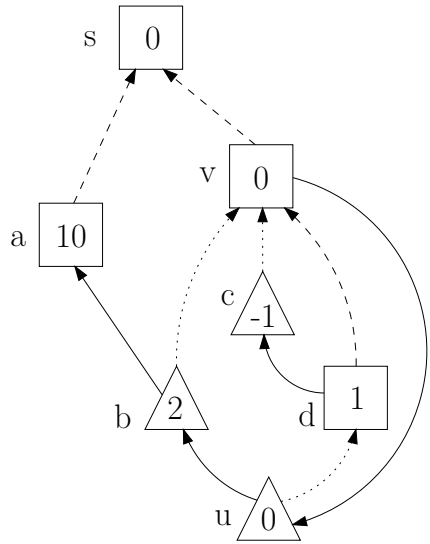}
  \end{center}
  \caption{A strategy tree.}
  \label{backedge-figure}
\end{figure}

We can now define our classification for profitable edges.
Let~$(v, u)$ be a profitable edge in the strategy~$\sigma$. We call this a
profitable \emph{back edge} if~$u$ is in $\subtree^{\sigma}(v)$, otherwise we call it a
profitable \emph{cross edge}. 

Figure~\ref{backedge-figure} gives an example of a strategy tree. In all of our
diagrams, dashed lines give a strategy $\sigma$ for player Max, and dotted lines
show Min's best response to the strategy of Max. The strategy tree contains
every vertex, and every edge that is either dashed or dotted. The subtree of~$v$
is the set~$\{v, b, c, d, u\}$. The edge $(v, u)$ is profitable because
$\val^{\sigma}(v) = 0$ and $\val^{\sigma}(u) = 1$. Since~$u$ is contained in the
subtree of~$v$, the edge $(v, u)$ is a profitable back edge.

%% file: backedges.tex
\section{Profitable Back Edges}
\label{pbe}

In this section we will expose the intimate connection between profitable back
edges and snares. We will show how every profitable back edge corresponds to
some snare that exists in the game. We will also define the concept of snare
consistency, and we will show how this concept is linked with the conditions
implied by Theorem~\ref{winningsnares}.

Our first task is to show how each profitable back edge corresponds to some Max
snare in the game. Recall that a Max snare consists of a set of vertices, and a
strategy for Max that is winning for the subgame induced by those vertices. We will
begin by defining the set of vertices for the snare that corresponds to a
profitable back edge. For a profitable back edge $(v, u)$ in a strategy~$\sigma$
we define the critical set, which is the vertices in $\subtree^{\sigma}(v)$ that
Min can reach when Max plays~$\sigma$. 

\begin{definition}[Critical Set]
If~$(v, u)$ is a profitable back edge in the strategy~$\sigma$, then we define
the critical set as $\critical^{\sigma}(v, u) = \{ w \in \subtree^\sigma(v)
\; : \;$ There is a path $\langle u, u_1, \dots u_k = w\rangle$ where for all
$i$ with $1 \le i \le k$ we have $u_i \in \subtree^\sigma(v)$ and if $u_i \in
V_\text{Max}$ then $u_{i+1} = \sigma(u_i) \}$.
\end{definition}

In the example given in Figure~\ref{backedge-figure}, the critical set of the
edge $(v, u)$ is $\{v, b, d, u\}$. The vertex~$b$ is in the critical set because
it is in the subtree of~$v$, and Min can reach it from~$u$ when Max plays
$\sigma$. In contrast, the vertex~$c$ is not in the critical set because
$\sigma(d) = v$, and therefore Min cannot reach~$c$ from~$u$ when Max plays
$\sigma$. The vertex~$a$ is not in the critical set because it is not in the
subtree of $v$. 

Note that in the example, $\sigma[v \mapsto u]$ is a winning strategy for the
subgame induced by critical set. The definition of the critical set is intended
to capture the largest connected subset of vertices contained in the subtree
of~$v$ for which $\sigma[v \mapsto u]$ is guaranteed to be a winning strategy.


\begin{proposition}
\label{besubgamewin}
Let~$(v, u)$ be a profitable back edge in the strategy~$\sigma$ and let $C$ be
$\critical^{\sigma}(v, u)$. The strategy $\sigma[v \mapsto u]$ is winning for
every vertex in $G \restriction C$.
\end{proposition}


We can now formally define the snare that is associated with each profitable
back edge that is encountered by strategy improvement. For a profitable back
edge $(v, u)$ in a strategy~$\sigma$ we define $\snare^{\sigma}(v, u) =
(\critical^{\sigma}(v, u), \chi)$ where $\chi(v) = \sigma[v \mapsto u](v)$ if $v
\in \critical^{\sigma}(v, u)$, and undefined at other vertices.
Proposition~\ref{besubgamewin} confirms that this meets the definition of a
snare.


We will now argue that the conditions given by Theorem~\ref{winningsnares} must be
observed in order for strategy improvement to terminate. We begin by defining a
concept that we call snare consistency. We say that a Max strategy is consistent
with a snare if Min's best response chooses an escape from that snare. 

\begin{definition}[Snare Consistency]
A strategy $\sigma$ is said to be consistent with the snare $(W, \chi)$ if
$\br(\sigma)$ uses some edge in $\escapes(W)$.
\end{definition}


In the example given in Figure~\ref{backedge-figure} we can see that $\sigma$ is
not consistent with $\snare^{\sigma}(v, u)$. This is because $\br(\sigma)$ does
not choose the edge $(b, a)$. However, once the edge $(v, u)$ is switched we can
prove that $\br(\sigma[v \mapsto u])$ must use the edge $(b, a)$. This is
because Min has no other way of connecting every vertex in
$\subtree^{\sigma}(v)$ to the sink, and if some vertex is not connected to the
sink then its valuation will rise to~$\infty$.
 
\begin{proposition}
\label{chooseescape}
Let~$(v, u)$ be a profitable back edge in the strategy $\sigma$. There is some
edge $(x, y)$ in $\escapes(\critical^{\sigma}(v, u))$ such that $\br(\sigma[v
\mapsto u])(x) = y$.
\end{proposition}


We can show that strategy improvement cannot terminate unless the current
strategy is consistent with every snare that exists in the game. This is because
every strategy that is not consistent with some snare must contain a profitable
edge. 

\begin{proposition}
\label{snareconsistency}
Let $\sigma$ be a strategy that is not consistent with a snare $(W, \chi)$.
There is a profitable edge $(v, u)$ in $\sigma$ such that $\chi(v) = u$.
\end{proposition}

These two propositions give us a new tool to study the process of strategy
improvement. Instead of viewing strategy improvement as a process that tries to
increase valuations, we can view it as a process that tries to force consistency
with Max snares. Proposition~\ref{snareconsistency} implies that this process
can only terminate when the current strategy is consistent with every Max snare
in the game. Therefore, the behaviour of strategy improvement on an example is
strongly related with the snares that exist for the strategy improver in that
example.



%% file: snares.tex
\section{Using Snares To Guide Strategy Improvement}
\label{snaresection}

In the previous sections, we have shown the strong link between snares and
strategy improvement. In this section we will show how this insight can be used
to guide strategy improvement. We will give a procedure that takes a strategy
that is inconsistent with some snare, and returns an improved strategy that is
consistent with that snare. Since the procedure is guaranteed to produce an
improved strategy, it can be used during strategy improvement as an alternative
to switching a profitable edge. We call algorithms that make use of this
procedure non-oblivious strategy improvement algorithms, and we give a
reasonable example of such an algorithm.

To define our procedure we will use Proposition~\ref{snareconsistency}. Recall
that this proposition implies that if a strategy $\sigma$ is inconsistent with a
snare $(W, \chi)$, then there is some profitable edge $(v, u)$ in $\sigma$ such
that $\chi(v) = u$. Our procedure will actually be a strategy improvement
switching policy. This policy will always choose to switch an edge that is
chosen by $\chi$ but not by the current strategy. As long as the current
strategy remains inconsistent with $(W, \chi)$ such an edge is guaranteed to
exist, and the policy terminates once the current strategy is consistent with
the snare. This procedure is shown as Algorithm~\ref{fixsnare}

\begin{algorithm}
\begin{algorithmic}
\WHILE{$\sigma$ is inconsistent with $(W, \chi)$}
\STATE $(v, w) :=$ Some edge where $\chi(v) = w$ and $(v, w)$ is profitable in
$\sigma$.
\STATE $\sigma := \sigma[v \mapsto u]$
\ENDWHILE
\STATE \textbf{return} $\sigma$
\end{algorithmic}
\caption{$\fixsnare(\sigma, (W, \chi))$}
\label{fixsnare}
\end{algorithm}

In each iteration the switching policy switches one vertex $v$ to an edge $(v,
u)$ with the property that $\chi(v) = u$, and it never switches a vertex at
which the current strategy agrees with $\chi$. It is therefore not difficult to
see that if the algorithm has not terminated after $|W|$ iterations then the
current strategy will agree with $\chi$ on every vertex in $W$. We can prove
that such a strategy must be consistent with $(W, \chi)$, and therefore the
switching policy must terminate after at most $|W|$ iterations.

\begin{proposition}
\label{fixsnareprop}
Let $\sigma$ be a strategy that is not consistent with a snare $(W, \chi)$.
Algorithm~\ref{fixsnare} will arrive at a strategy $\sigma'$ which is consistent
with $(W, \chi)$ after at most $|W|$ iterations.
\end{proposition}

Since $\fixsnare$ is implemented as a strategy improvement switching policy that
switches only profitable edges, the strategy that is produced must be an improved
strategy. Therefore, at any point during the execution of strategy improvement
we can choose not to switch a subset of profitable edges and run $\fixsnare$
instead. Note that the strategy produced by $\fixsnare$ may not be reachable
from the current strategy by switching a subset of profitable edges. This is
because $\fixsnare$ switches a sequence of profitable edges, some of which may
not have been profitable in the original strategy. 
 
We propose a new class of strategy improvement algorithms that are aware of
snares. These algorithms will record a snare for every profitable back edge that
they encounter during their execution. In each iteration these algorithms can
either switch a subset of profitable edges or run the procedure $\fixsnare$ on
some recorded snare that the current strategy is inconsistent with. We call
these algorithms non-oblivious strategy improvement algorithms, and the general
schema that these algorithms follow is shown in Algorithm~\ref{nonobliv}.

\begin{algorithm}
\begin{algorithmic}
\STATE $S := \emptyset$
\WHILE{$\sigma$ has a profitable edge}
\STATE $S := S \cup \{\snare^{\sigma}(v, u) \; : \; (v, u) \text{ is a
profitable back edge in $\sigma$} \}$ 
\STATE $\sigma := \policy(\sigma, S)$
\ENDWHILE
\STATE \textbf{return} $\sigma$
\end{algorithmic}
\caption{$\nonobliv(\sigma)$}
\label{nonobliv}
\end{algorithm}

Recall that oblivious strategy improvement algorithms required a switching
policy to specify which profitable edges should be switched in each iteration.
Clearly, non-oblivious strategy improvement algorithms require a similar method
to decide whether to apply the procedure $\fixsnare$ or to pick some subset of
profitable edges to switch. Moreover, they must decide which snare should be
used when the procedure $\fixsnare$ is applied. We do not claim to have the
definitive non-oblivious switching policy, but in the rest of this section we
will present one reasonable method of constructing a non-oblivious version of an
oblivious switching policy. We will later show that our non-oblivious strategy
improvement algorithms behave well on the examples that are known to cause
exponential time behaviour for oblivious strategy improvement.

We intend to take an oblivious switching policy~$\alpha$ as the base of our
non-oblivious switching policy. This means that when we do not choose to use
the procedure $\fixsnare$, we will switch the subset of profitable edges that
would be chosen by~$\alpha$. Our goal is to only use $\fixsnare$ when doing so
is guaranteed to yield a larger increase in valuation than applying~$\alpha$.
Clearly, in order to achieve this we must know how much the valuations increase
when~$\alpha$ is applied and how much the valuations increase when $\fixsnare$
is applied.

Determining the increase in valuation that is produced by applying an oblivious
switching policy is easy. Since every iteration of oblivious strategy
improvement takes polynomial time, We can simply switch the edges and measure
the difference between the current strategy and the one that would be produced.
Let $\sigma$ be a strategy and let $P$ be the set of edges that are profitable
in $\sigma$. For an oblivious switching policy $\alpha$ the increase of applying
$\alpha$ is defined to be:
\begin{equation*}
\increase(\alpha, \sigma) = \sum_{v \in V} (\val^{\sigma[\alpha(P)]}(v) -
\val^{\sigma}(v))
\end{equation*}

We now give a lower bound on the increase in valuation that an application of
$\fixsnare$ produces. Let $(W, \chi)$ be a snare and suppose that the current
strategy $\sigma$ is inconsistent with this snare. Our lower bound is based on
the fact that $\fixsnare$ will produce a strategy that is consistent with the 
snare. This means that Min's best response is not currently choosing an escape
from the snare, but it will be forced to do so after $\fixsnare$ has been
applied. It is easy to see that forcing the best response to use a different
edge will cause an increase in valuation, since otherwise the best response
would already be using that edge. Therefore, we can use the increase in
valuation that will be obtained when Min is forced to use and escape. We define:
\begin{equation*}
\snareincrease^\sigma(W, \chi) = \min\{(\val^{\sigma}(y) + \weight(x)) -
\val^{\sigma}(x) \; : \; (x, y) \in \escapes(W)\}
\end{equation*}
This expression gives the smallest possible increase in valuation that can
happen when Min is forced to use an edge in $\escapes(W)$. We can prove that
applying $\fixsnare$ will cause an increase in valuation of at least this
amount.

\begin{proposition}
\label{lowersnare}
Let~$\sigma$ be a strategy that is not consistent with a snare~$(W, \chi)$, and
let~$\sigma'$ be the result of $\fixsnare(\sigma, (W, \chi))$. We have:
\begin{equation*}
\sum_{v \in V} (\val^{\sigma'}(v) - \val^{\sigma}(v)) \ge
\snareincrease^\sigma(W, \chi)
\end{equation*}
\end{proposition}

We now have the tools necessary to construct our proposed augmentation scheme,
which is shown as Algorithm~\ref{nonobv}. The idea is to compare the increase
obtained by applying $\alpha$ and the increase obtained by applying $\fixsnare$
with the best snare that has been previously recorded, and then to only apply
$\fixsnare$ when it is guaranteed to yield a larger increase in valuation.

\begin{algorithm}
\begin{algorithmic}
\STATE $(W, \chi) := \argmax_{(X, \mu) \in S} \snareincrease^{\sigma}(X, \mu)$
\IF{$\increase(\alpha, \sigma) > \snareincrease^{\sigma}(W, \chi)$}
\STATE $P := \{ (v, u) \; : \; (v, u) \text{ is profitable in } \sigma \}$
\STATE $\sigma := \sigma[\alpha(P)]$
\ELSE
\STATE $\sigma := \fixsnare(\sigma, (W, \chi))$
\ENDIF
\STATE \textbf{return} $\sigma$
\end{algorithmic}
\caption{$(\augment(\alpha))(\sigma, S)$}
\label{nonobv}
\end{algorithm}



%% file: friedmann.tex
\section{Comparison With Oblivious Strategy Improvement}
\label{friedmann}

In this section we will demonstrate how non-oblivious strategy improvement can
behave well in situations where oblivious strategy improvement has exponential
time behaviour. Unfortunately, there is only one source of examples with such
properties in the literature, and that is the family of examples given by
Friedmann. In fact, Friedmann gives two slightly different families of hard
examples. The first type is the family that that forces exponential behaviour
for the all-switches policy~\cite{Fri09}, and the second type is the family that
forces exponential behaviour for both all-switches and optimal switching
policies~\cite{Fri09preprint}. Although our algorithm performs well on both
families, we will focus on the example that was designed for optimal switching
policies because it is the most interesting of the two.

This section is split into two parts. In the first half of this section we will
study a component part of Friedmann's example upon which the procedure
$\fixsnare$ can out perform an optimal switching policy. 
This implies that there are situations in which our augmentation scheme will
choose to use $\fixsnare$. In the second half, we will show how the good
performance on the component part is the key property that allows our
non-oblivious strategy improvement algorithms to terminate quickly on
Friedmann's examples.


\subsection{Optimal Switching Policies}

\begin{figure}
  \begin{center}
    \includegraphics[width=0.35\textwidth]{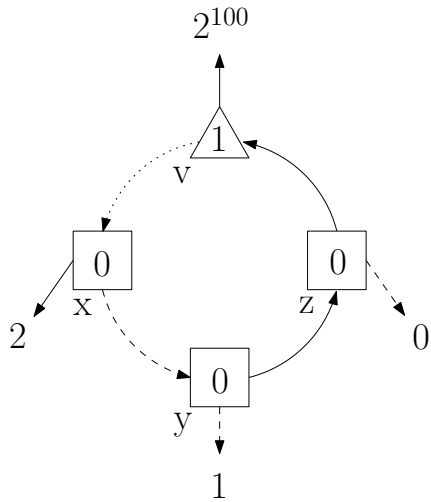}
  \end{center}
  \caption{A component of Friedmann's exponential time example.}
  \label{schewecycle}
\end{figure}

We have claimed that the procedure $\fixsnare$ can cause a greater increase in
valuation than switching any subset of profitable edges. We will now give an
example upon which this property holds. The example that we will consider is
shown in Figure~\ref{schewecycle}, and it is one of the component parts of
Friedmann's family of examples that force optimal policies to take an
exponential number of steps~\cite{Fri09preprint}. 

The diagram shows a strategy for Max as a set of dashed edges. It also shows
Min's best response to this strategy as a dotted edge. Even though this example
could be embedded in an arbitrary game, we can reason about the behaviour of
strategy improvement by specifying, for each edge that leaves the example, the
valuation of the successor vertex that the edge leads to. These valuations are
shown as numbers at the end of each edge that leaves the example.

In order to understand how strategy improvement behaves we must determine the
set of edges that are profitable for our strategy. There are two edges that are
profitable: the edge~$(z, v)$ is profitable because the valuation of~$v$ is~$2$
which is greater than~$0$, and the edge at~$x$ that leaves the example is
profitable because leaving the example gives a valuation of~$2$ and the
valuation of~$y$ is~$1$. The edge~$(y, z)$ is not profitable because the
valuation of~$z$ is~$0$, which is smaller than the valuation of~$1$ obtained by
leaving the example at~$y$.

For the purposes of demonstration, we will assume that no other edge is
profitable in the game into which the example is embedded. Furthermore, we will
assume that no matter what profitable edges are chosen to be switched, the
valuation of every vertex not contained in the example will remain constant.
Therefore, the all-switches policy will switch the edges $(z, v)$ and the edge
leading away from the example at the vertex $x$. It can easily be verified that
this is also the optimal subset of profitable edges, and so the all-switches and
the optimal policies make the same decisions for this strategy. After switching the
edges chosen by the two policies, the valuation of $x$ will rise to $2$, the
valuation of $z$ will rise to $3$, and the valuation of $y$ remain at $1$.



By contrast, we will now argue that non-oblivious strategy improvement would
raise the valuations of $x$, $y$, and $z$ to $2^{100} + 1$. Firstly, it is
critical to note that the example is a snare. If we set $W = \{v, x, y, z\}$ and
choose $\chi$ to be the partial strategy for Max that chooses the edges $(x,
y)$, $(y, z)$, and $(z, v)$, then $(W, \chi)$ will be a snare in every game into
which the example is embedded. This is because there is only one cycle in the
subgame induced by $W$ when Max plays $\chi$, and this cycle has positive
weight. 

Now, if the non-oblivious strategy improvement algorithm was aware of the snare
$(W, \chi)$ then the lower bound given by Proposition~\ref{lowersnare} would be
$2^{100}$. This is because closing the cycle forces Min's best response to
use escape edge to avoid losing the game. Since $2^{100}$ is much larger than
the increase obtained by the optimal switching policy, the policies
$\augment(\all)$ and $\augment(\schewe)$ will choose to run $\fixsnare$ on the
snare $(W, \chi)$. Once consequence of this is that the policy $\schewe$ is no
longer optimal in the non-oblivious setting.

\subsection{Friedmann's Exponential Time Examples}

The example that we gave in the previous subsection may appear to be trivial. After
all, if the valuations outside the example remain constant then both the
all-switches and optimal switching policies will close the cycle in two
iterations. A problem arises, however, when the valuations can change. Note that
when we applied the oblivious policies to the example, no progress was made towards
closing the cycle. We started with a strategy that chose to close the cycle at
only one vertex, and we produced a strategy that chose to close the cycle at
only one vertex. When the assumption that valuations outside the example are
constant is removed, it becomes possible for a well designed game to delay the
closing of the cycle for an arbitrarily large number of iterations simply by
repeating the pattern of valuations that is shown in Figure~\ref{schewecycle}.

\begin{figure}
  \begin{center}
    \includegraphics[width=0.7\textwidth]{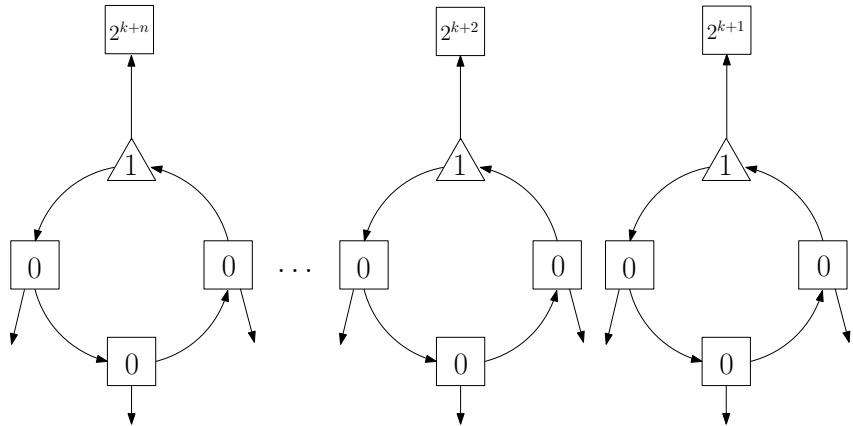}
  \end{center}
  \caption{The bits of a binary counter.}
  \label{bits}
\end{figure}

Friedmann's family of examples exploits this property to build a binary counter,
which uses the subgame shown in Figure~\ref{schewecycle} to represent the
bits. The general idea of this approach is shown in Figure~\ref{bits}.
Friedmann's example uses $n$ instances of the cycle, indexed~$1$ through~$n$.
These bits are interconnected in a way that enforces two properties on both the
all-switches and the optimal switching policies. Firstly, the ability to prevent
a cycle from closing that we have described is used to ensure that the cycle
with index~$i$ can only be closed after every cycle with index smaller
than~$i$ has been closed. Secondly, when the cycle with index~$i$ is closed,
every cycle with index smaller than~$i$ is forced to open. Finally, every cycle
is closed in the optimal strategy for the example. Now, if the initial strategy
is chosen so that every cycle is open, then these three properties are
sufficient to force both switching policies to take at least $2^n$ steps before
terminating.
 
The example works by forcing the oblivious switching policy to make the same
mistakes repeatedly. To see this, consider the cycle with index $n-1$. When the
cycle with index $n$ is closed for the first time, this cycle is forced open.
The oblivious optimal switching policy will then not close it again for at least
another $2^{n-1}$ steps. By contrast, the policies $\augment(\all)$ and
$\augment(\schewe)$ would close the cycle again after a single iteration. This
breaks the exponential time behaviour, and it turns out that both of our
policies terminate in polynomial time on Friedmann's examples.

Of course, for Friedmann's examples we can tell simply by inspection that Max
always wants to keep the cycle closed. It is not difficult, however, to imagine
an example which replaces the four vertex cycle with a complicated subgame, for
which Max had a winning strategy and Min's only escape is to play to the vertex
with a large weight. This would still be a snare, but the fact that it is a
snare would only become apparent during the execution of strategy improvement.
Nevertheless, as long as the complicated subgame can be solved in polynomial
time by non-oblivious strategy improvement, the whole game will also be solved
in polynomial time. This holds for exactly the same reason as the polynomial
behaviour on Friedmann's examples: once the snare representing the subgame has
been recorded then consistency with that snare can easily be enforced in the
future.

%% file: conclusions.tex
\section{Conclusions and Further Work}

In this paper we have uncovered and formalized a strong link between the snares
that exist in a game and the behaviour of strategy improvement on that game. We
have shown how awareness of this link can be used to guide the process of
strategy improvement. With our augmentation procedure we gave one reasonable
method of incorporating non-oblivious techniques into traditional strategy
improvement, and we have demonstrated how these techniques give rise to good
behaviour on the known exponential time examples.

It must be stressed that we are not claiming that simply terminating in
polynomial time on Friedmann's examples is a major step forward. After all, the
randomized switching policies of Bj\"orklund and Vorobyov~\cite{bv07} have the
same property. What is important is that our strategy improvement algorithms are
polynomial because they have a better understanding of the underlying structure
of strategy improvement. Friedmann's examples provide an excellent cautionary
tale that shows how ignorance of this underlying structure can lead to
exponential time behaviour.

There are a wide variety of questions that are raised by this work. Firstly, we
have the structure of snares in parity and mean-payoff games.
Theorem~\ref{winningsnares} implies that all algorithms that find winning
strategies for parity and mean payoff games must, at least implicitly, consider
snares. We therefore propose that a thorough and complete understanding of how
snares arise in a game is a necessary condition for devising a polynomial time
algorithm for these games.

It is not currently clear how the snares in a game affect the difficulty of
solving that game. It is not difficult, for example, to construct a game in
which there an exponential number of Max snares: in a game in which every weight
is positive there will be a snare for every connected subset of vertices.
However, games with only positive weights have been shown to be very easy to
solve~\cite{kgz06}. Clearly, the first challenge is to give a clear formulation
of how the structure of the snares in a given game affects the difficulty of
solving it.

In our attempts to construct intelligent non-oblivious strategy improvement
algorithms we have continually had problems with examples in which Max and Min
snares overlap. By this we mean that the set of vertices that define the
subgames of the snares have a non empty intersection. We therefore think that
studying how complex the overlapping of snares can be in a game may lead to
further insight. There are reasons to believe that these overlappings cannot be
totally arbitrary, since they arise from the structure of the game graph and the
weights assigned to the vertices.

We have presented a non-oblivious strategy improvement algorithm that passively
records the snares that are discovered by an oblivious switching policy, and
then uses those snares when doing so is guaranteed to lead to a larger increase
in valuations. While we have shown that this approach can clearly outperform
traditional strategy improvement, it does not appear to immediately lead to a
proof of polynomial time termination. It would be interesting to find an
exponential time example for the augmented versions of the all-switches policy
or of the optimal policy. This may be significantly more difficult since it
is no longer possible to trick strategy improvement into making slow progress by
forcing it to repeatedly close a small number of snares. 

There is no inherent reason why strategy improvement algorithms should be
obsessed with trying to increase valuations as much as possible in each
iteration. Friedmann's exponential time example for the optimal policy
demonstrates that doing so in no way guarantees that the algorithm will always
make good progress. Our work uncovers an alternate objective that strategy
improvement algorithms can use to measure their progress. Strategy improvement
algorithms could actively try to discover the snares that exist in the game, or
they could try and maintain consistency with as many snares as possible, for
example. There is much scope for an intelligent snare based strategy improvement
algorithm.

We have had some limited success in designing intelligent snare based strategy
improvement algorithms for parity games. We have developed a non-oblivious
strategy improvement algorithm which, when given a list of known snares in the
game, either solves the game or finds a snare that is not in the list of known
snares. This gives the rather weak result of a strategy improvement algorithm
whose running time is polynomial in $|V|$ and $k$, where $k$ is the number of
Max snares that exist in the game. This is clearly unsatisfactory since we have
already argued that $k$ could be exponential in the number of vertices. However,
this is one example of how snares can be applied to obtain new bounds for
strategy improvement. As an aside, the techniques that we used to obtain this
algorithm do not generalize to mean-payoff games. Finding a way to accomplish
this task for mean-payoff games is an obvious starting point for designing
intelligent snare based algorithms for this type of game. \\

\noindent\textbf{Acknowledgements.} I am indebted to Marcin Jurdzi\'nski for his
guidance, support, and encouragement during the preparation of this paper.

%% file: appendix.tex
\newpage

\section{Proofs for Section \ref{snareintro}}

\subsection{Proof of Theorem \ref{winningsnares}}

\begin{proof}
For the sake of contradiction, suppose that $\tau$ is a winning strategy for
$S$ that does not choose an edge in $\escapes(W)$. Since $\chi$ also
does not choose an edge that leaves $W$, we have that $\play(v, \chi, \tau)$
never leaves the set $W$, for every vertex $v$ in $W$. Furthermore, since $\chi$
is a winning strategy for the subgame induced by $W$ we have
$\meanpayoff(\play(v, \chi, \tau)) > 0$ for every vertex $v$ in $W$, which
contradicts the fact that $\tau$ is a winning strategy for $S$. 
\end{proof}

\section{Proofs for Section \ref{strategytrees}}

\subsection{Proof of Proposition \ref{poscycle}}

\begin{proof}
The algorithm is shown as Algorithm~\ref{zeromean}. We use the notation $G
\restriction U$ to refer to the sub-game of $G$ induced by the set of vertices
$U$. Its correctness follows from a result of Zielonka~\cite{zielonka98} which
was originally shown for parity games, but identical techniques apply in this
setting. Let $W$ be a set of vertices, we define the set of vertices from which
Max can force the token into $W$ in one step as
\begin{equation*}
\begin{split}
\pre(W) = \{v \in V_{\text{Max}} \; : \; \text{There is an edge} (v, u) \text{
with } u \in W\} \\\cup \{v \in V_{\text{Min}} \; : \; \text{All edges} (v, u) \text{
have } u \in W\}.
\end{split}
\end{equation*}
We then define the attractor of $W$ to be the set of vertices from which Max can
force play into $W$.
\begin{align*}
W_0 &= W\\
W_i &= W_{i-1} \cup \pre(W_{i-1})\\
\attr(W) &= W_{|V|}
\end{align*}
Zielonka showed that if $W$ is a subset of Max's winning set, which is
the set of vertices with value greater than~0, then both winning sets can be
found by solving the sub-game $G \restriction \attr(W)$.

In our setting the algorithm $\alpha$ finds the set $W$, and it is clear that
the loop computes $\attr(W)$. Therefore, we get that our algorithm finds the
0-mean partition. Moreover, since each recursive call decreases the size of the
game by at least one vertex we get that at most $|V|$ calls to $\alpha$ are
made. 

\begin{algorithm}
\begin{algorithmic}
\STATE $\sigma := \alpha(\sigma)$
\WHILE{There is an edge $(v, u)$ with $\val^{\sigma}(v) < \infty$ and
$\val^{\sigma}(u) = \infty$}
\STATE $\sigma := \sigma[v \mapsto u]$
\ENDWHILE
\STATE $W_{>0} := \{v \; : \; \val^{\sigma}(v) = \infty\}$
\STATE $U := V \setminus W_{>0}$
\STATE $(W_{\le0}', W_{>0}') := \zeromean(\sigma, \alpha, G \restriction U)$
\STATE \textbf{return} $(W_{\le0}', W_{>0} \cup W_{>0}')$
\end{algorithmic}
\caption{$\zeromean(\sigma, \alpha, G)$}
\label{zeromean}
\end{algorithm}
\end{proof}

\section{Proofs for Section \ref{pbe}}

\subsection{Proof of Proposition \ref{besubgamewin}}

\begin{proof}
Since $C$ is a critical set it must be the case that every vertex in $C$ must be
in the subtree of $v$ according to $\sigma$, and this implies that $\sigma[v
\mapsto u](w)$ is not the sink for every vertex $w$ in $C$. Note that only paths
ending at the sink can have finite valuations, and that no such paths can exist
when $\sigma[v \mapsto u]$ is played in $G \restriction C$. This implies that
$\val^{\sigma[v \mapsto u]}_{G \restriction C}(w)$ is either $\infty$ or
$-\infty$, and we will argue that the latter is not possible.

Suppose for the sake of contradiction that there is a vertex $w$ with the
property $\val^{\sigma[v \mapsto u]}_{G \restriction C}(w) = -\infty$. Let
$\tau$ be $\br_{G \restriction C}(\sigma[v \mapsto u])$. We define $\tau'$ to be
a strategy $G$ that follows $\tau$ on the vertices in $G \restriction C$ and
makes arbitrary decisions at the other vertices. For every vertex $w$ in
$V_{\text{Min}}$ we choose some edge $(w, x)$ and define
\begin{equation*}
\tau'(w) = \begin{cases}
\tau(w) & \text{if } w \in C, \\
x & \text{otherwise.}
\end{cases}
\end{equation*}
Now consider $\sigma[v \mapsto u]$ played against $\tau'$ on the game $G$. Note
that neither of the two strategies choose an edge that leaves the set $C$ and so
$\play_G(w, \sigma[v \mapsto u], \tau') = \play_{G \restriction C}(w, \sigma[v
\mapsto u], \tau')$ for every vertex $w$ in $C$. Since valuations can be derived
from the play, this implies that $\val^{\sigma[v \mapsto u], \tau'}_G(w) =
-\infty$. By the properties of the best response we have for every vertex $w$ in
$C$.
\begin{equation*}
\val^{\sigma[v \mapsto u]}_G(w) \le \val^{\sigma[v \mapsto u], \tau'}_G(w) = -\infty
< \val^{\sigma}_G(w)
\end{equation*}
This contradicts Theorem~\ref{bv1}, and so we can conclude that $\val^{\sigma[v
\mapsto u]}_{G \restriction C}(w) = \infty$ for every vertex $w$ in $C$. 
\end{proof}




\subsection{Proof of Proposition \ref{chooseescape}}

\begin{proof}
Consider a strategy $\tau$ for player Min for which there is no
edge $(x, y)$ in $\escapes(\critical^{\sigma}(v, u))$ with $\tau(x) = y$. We
argue that $\val^{\sigma[v \mapsto u], \tau}(w) = \infty$ for every vertex
$w$ in $\critical^{\sigma}(v, u)$. Note that neither $\sigma[v \mapsto u]$ or
$\tau$ chooses an edge that leaves $\critical^{\sigma}(v, u)$, which implies
that $\play(w, \sigma[v \mapsto u], \tau)$ does not leave $\critical^{\sigma}(v,
u)$, for every vertex $w$ in $\critical^{\sigma}(v, u)$. By
Proposition~\ref{besubgamewin} we have that $\sigma[v \mapsto u]$ is a winning
strategy for $G \restriction \critical^{\sigma}(v, u)$, and therefore
$\val^{\sigma[v \mapsto u], \tau}(w) = \infty$ for every vertex $w$ in
$\critical^{\sigma}(v, u)$.

We will now construct a strategy for Min which, when played against $\sigma[v
\mapsto u]$, guarantees a finite valuation for some vertex in
$\critical^{\sigma}(v, u)$. Let $(x, y)$ be some edge in
$\escapes(\critical^{\sigma}(v, u))$. We define the Min strategy $\tau$, for
every vertex $w$ in $V_\text{Min}$ as
\begin{equation*}
\tau(w) = \begin{cases}
y & \text{if } w = x, \\
\br(\sigma)(w) & \text{otherwise.}
\end{cases}
\end{equation*}
By definition of critical set we have that $y$ cannot be in the subtree of $v$,
since otherwise it would also be in $\critical^{\sigma}(v, u)$. This implies
that $\play(y, \sigma, \br(\sigma)) = \play(y, \sigma[v \mapsto u], \tau)$,
since $\tau = \br(\sigma)$ on every vertex that is not in
$\subtree^{\sigma}(v)$, and $\sigma = \sigma[v \mapsto u]$ on every vertex that
is not $v$. From this we can conclude that $\val^{\sigma[v \mapsto u], \tau}(y)
= \val^{\sigma}(y) < \infty$. By construction of $\tau$ we have that $\val^{\sigma[v \mapsto u],
\tau}(x) = \val^{\sigma[v \mapsto u], \tau}(y) + \weight(x)$, and so we also
have $\val^{\sigma[v \mapsto u], \tau}(x) < \infty$.

In summary, we have shown that every Min strategy $\tau$ that does not use an
edge in $\escapes(\critical^{\sigma}(v, u))$ has the property $\val^{\sigma[v
\mapsto u], \tau}(w) = \infty$ for every vertex $v$ in
$\critical^{\sigma}(v, u)$. We have also shown that there is a Min
strategy $\tau$ which guarantees $\val^{\sigma[v \mapsto u], \tau}(w) < \infty$
for some vertex $w$ in $\critical^{\sigma}(v, u)$. From the properties of a best
response we can conclude that Min must use some edge in
$\escapes(\critical^{\sigma}(v, u))$. 
\end{proof}

\subsection{Proof of Proposition~\ref{snareconsistency}}
\begin{proof}
In order to prove the claim we will construct an alternate game. We
define the game $G' = (V, V_\text{Max}, V_\text{Min}, E', \weight)$ where:
\begin{equation*}
E' = \{(v, u) \; : \; \sigma(v) = u \text{ or } \br_G(\sigma)(v) = u \text{ or }
\chi(v) = u \}.
\end{equation*}
In other words, we construct a game where Min is forced to play
$\br_G(\sigma)(v)$ and Max's strategy can be constructed using a combination of
the edges used by $\sigma$ and $\chi$. Since Min is forced to play
$\br_G(\sigma)(v)$ we have that $\val^{\sigma}_G(v) = \val^{\sigma}_{G'}(v)$ for
every vertex $v$. To decide if an edge is profitable we compare two valuations,
and since the valuation of $\sigma$ is the same in both $G$ and $G'$ we have
that an edge is profitable for $\sigma$ in $G$ if and only if it is profitable
for $\sigma$ in $G'$. Note also that the only way $\sigma$ can be modified in
$G'$ is to choose an edge that is chosen by $\chi$ but not by $\sigma$.
Therefore, to prove our claim it is sufficient to show that $\sigma$ has a
profitable edge in $G'$.

We define the strategy:
\begin{equation*}
\chi'(v) = \begin{cases}
\chi(v) & \text{if } v \in W, \\
\sigma(v) & \text{otherwise.}
\end{cases}
\end{equation*}
We will argue that~$\chi'$ is a better strategy than~$\sigma$ in~$G'$. The
definition of a snare implies that~$\chi$ is a winning strategy for the sub-game
induced by~$W$, and by assumption we have
that~$\br(\sigma)$ does not use an edge in~$\escapes(W)$. We therefore have that
$\val^{\chi'}_{G'}(v) = \infty$ for every vertex~$v$ in~$W$. On the other hand,
since we are considering the positive cycle problem, we know that
$\val^{\sigma}_{G'}(v) < \infty$ for every vertex~$v$ in~$W$.  This implies
that~$\sigma$ is not the optimal strategy in~$G'$.  Theorem~\ref{bv1} implies
that all non-optimal strategies must have at least one profitable edge, and the
only edges that can be profitable in~$G'$ are those chosen by~$\chi$.
Therefore there is some edge chosen by~$\chi$ that is profitable for~$\sigma$
in~$G'$ and as we have argued this also means that the edge is profitable
for~$\sigma$ in~$G$. 
\end{proof}

\section{Proofs for Section \ref{snaresection}}
\subsection{Proof of Proposition~\ref{fixsnareprop}}

\begin{proof}
By Proposition~\ref{snareconsistency} we know that as long as the current
strategy is not consistent with the snare $(W, \chi)$ there must be an edge~$(v,
u)$ with $\chi(v) = u$ that is profitable in~$\sigma$. The switching policy will
always choose this edge, and will terminate once the current strategy is
consistent with the snare. Therefore in each iteration the number of vertices
upon which~$\sigma$ and~$\chi$ differ decreases by~$1$.  It follows that after
at most $|W|$ iterations we will have $\sigma(v) = \chi(v)$ for every vertex~$v$
in~$W$. Since~$\chi$ is a winning strategy for the sub-game induced by~$W$ we
have that player Min must choose some edge that leaves~$W$ to avoid losing once
this strategy has been reached. 
\end{proof}

\subsection{Proof of Proposition \ref{lowersnare}}
\begin{proof}
We will prove this proposition by showing that there exists some vertex $w$ with
the property $\val^{\sigma'}(w) - \val^{\sigma}(w) \ge \snareincrease(W, \chi)$.
Since the procedure $\fixsnare$ switches only profitable edges we have by
Theorem~\ref{bv1} that $\val^{\sigma'}(v) - \val^{\sigma}(v) \ge 0$ for every
vertex $v$. Therefore,
this is sufficient to prove the proposition because $\sum_{v \in
V}(\val^{\sigma'}(v) - \val^{\sigma}(v)) \ge \val^{\sigma'}(w) -
\val^{\sigma}(w)$.

Proposition~\ref{fixsnareprop} implies that $\sigma'$ is consistent with the
snare $(W, \chi)$. By the definition of snare consistency, this implies that
$\br(\sigma')$ must use some edge $(w, x)$ in $\escapes(W)$. We therefore have that
$\val^{\sigma'}(w) = \val^{\sigma'}(x) + \weight(w)$.  Since the $\fixsnare$
procedure switches only profitable edges, we have by Theorem~\ref{bv1} that
$\val^{\sigma'}(x) \ge \val^{\sigma}(x)$. The increase at $x$ is
therefore
\begin{align*}
\val^{\sigma'}(w) - \val^{\sigma}(w) &= \val^{\sigma' }(x) + \weight(w) -
\val^{\sigma}(w)\\
&\ge \val^{\sigma}(x) + \weight(w) - \val^{\sigma}(w) \\
&\ge \snareincrease(W, \chi) 
\end{align*}
\end{proof}

%% file: paper.bbl
\begin{thebibliography}{10}

\bibitem{bsv03}
H.~Bj\"{o}rklund, S.~Sandberg, and S.~Vorobyov.
\newblock A discrete subexponential algorithm for parity games.
\newblock In {\em Proceedings of the 20th Annual Symposium on Theoretical
  Aspects of Computer Science}, volume 2607 of {\em LNCS}, pages 663--674,
  London, UK, 2003. Springer-Verlag.

\bibitem{bv07}
H.~Bj\"{o}rklund and S.~Vorobyov.
\newblock A combinatorial strongly subexponential strategy improvement
  algorithm for mean payoff games.
\newblock {\em Discrete Applied Mathematics}, 155(2):210--229, 2007.

\bibitem{condon93}
A.~Condon.
\newblock On algorithms for simple stochastic games.
\newblock In J.-Y. Cai, editor, {\em Advances in Computational Complexity
  Theory, volume 13 of DIMACS Series in Discrete Mathematics and Theoretical
  Computer Science}, pages 51--73. American Mathematical Society, 1993.

\bibitem{EmersonJutla93}
E.~A. Emerson, C.~S. Jutla, and A.~P. Sistla.
\newblock On model-checking for fragments of $\mu$-calculus.
\newblock In C.~Courcoubetis, editor, {\em Computer Aided Verification, 5th
  International Conference, CAV'93}, volume 697 of {\em LNCS}, pages 385--396.
  Springer-Verlag, 1993.

\bibitem{Fri09}
O.~Friedman.
\newblock A super-polynomial lower bound for the parity game strategy
  improvement algorithm as we know it.
\newblock In {\em Logic in Computer Science (LICS)}. IEEE, 2009.

\bibitem{Fri09preprint}
O.~Friedman.
\newblock A super-polynomial lower bound for the parity game strategy
  improvement algorithm as we know it.
\newblock Preprint, January 2009.

\bibitem{howard60}
R.~Howard.
\newblock {\em Dynamic Programming and Markov Processes}.
\newblock Technology Press and Wiley, 1960.

\bibitem{jpz06}
M.~Jurdzi{\'n}ski, M.~Paterson, and U.~Zwick.
\newblock A deterministic subexponential algorithm for solving parity games.
\newblock In {\em Proceedings of {ACM}-{SIAM} Symposium on Discrete Algorithms,
  SODA 2006}, pages 117--123. ACM/SIAM, 2006.

\bibitem{kgz06}
L.~Khachiyan, V.~Gurvich, and J.~Zhao.
\newblock Extending dijkstra’s algorithm to maximize the shortest path by
  node-wise limited arc interdiction.
\newblock In {\em Computer Science -- Theory and Applications}, volume 3967 of
  {\em LNCS}, pages 221--234. Springer, 2006.

\bibitem{ll69}
T.~M. Liggett and S.~A. Lippman.
\newblock Stochastic games with perfect information and time average payoff.
\newblock {\em SIAM Review}, 11(4):604--607, 1969.

\bibitem{mansoursingh99}
Y.~Mansour and S.~P. Singh.
\newblock On the complexity of policy iteration.
\newblock In K.~B. Laskey and H.~Prade, editors, {\em UAI '99: Proceedings of
  the Fifteenth Conference on Uncertainty in Artificial Intelligence}, pages
  401--408. Morgan Kaufmann, 1999.

\bibitem{puri95}
A.~Puri.
\newblock {\em Theory of Hybrid Systems and Discrete Event Systems}.
\newblock PhD thesis, University of California, Berkeley, 1995.

\bibitem{schewe08}
S.~Schewe.
\newblock An optimal strategy improvement algorithm for solving parity and
  payoff games.
\newblock In {\em Computer Science Logic}, volume 5213 of {\em LNCS}, pages
  369--384. Springer, 2008.

\bibitem{stirling95}
C.~Stirling.
\newblock Local model checking games (extended abstract).
\newblock In I.~Lee and S.~A. Smolka, editors, {\em CONCUR'95: Concurrency
  Theory, 6th International Conference}, volume 962 of {\em LNCS}, pages 1--11.
  Springer-Verlag, 1995.

\bibitem{jurdzinski00b}
J.~V{\"{o}}ge and M.~Jurdzi{\'{n}}ski.
\newblock A discrete strategy improvement algorithm for solving parity games
  ({E}xtended abstract).
\newblock In E.~A. Emerson and A.~P. Sistla, editors, {\em Computer Aided
  Verification, 12th International Conference, CAV 2000, Proceedings}, volume
  1855 of {\em LNCS}, pages 202--215, Chicago, IL, USA, 2000. Springer-Verlag.

\bibitem{zielonka98}
W.~Zielonka.
\newblock Infinite games on finitely coloured graphs with applications to
  automata on infinite trees.
\newblock {\em Theoretical Computer Science}, 200:135--183, 1998.

\bibitem{patersonzwick96}
U.~Zwick and M.~Paterson.
\newblock The complexity of mean payoff games on graphs.
\newblock {\em Theoretical Computer Science}, 158(1--2):343--359, 1996.

\end{thebibliography}
